\documentclass[11pt,a4]{article}

\usepackage{a4wide}
\usepackage[latin1]{inputenc}
\usepackage[dvips]{graphicx}
\usepackage{color}
\usepackage{amsfonts}
\usepackage{multirow}

\newcommand{\qed}{\hfill$\Box$ \vspace{0.5 cm}}
\newenvironment{proof}{\noindent {\it Proof~: }}{\qed}

\newcommand{\point}{\textrm{\Huge{\hspace{-1pt}.}}}
\newtheorem{property}{Property}
\newtheorem{definition}{Definition}
\newtheorem{theorem}{Theorem}
\newtheorem{lemma}{Lemma}

\title{Computing communities in large networks using random walks}
\author{Pascal Pons and Matthieu Latapy}

\begin{document}

\begin{center}
{\Large{\bf Computing communities in large networks using random walks}}\\
\vspace{0.5cm}
{\Large Pascal Pons and Matthieu Latapy}\\ \vspace{2mm}
LIAFA -- CNRS and University Paris 7 -- 2 place Jussieu, 75251 Paris Cedex 05, France\\
(pons, latapy)@liafa.jussieu.fr
\end{center}

\abstract{ 
{\footnotesize Dense subgraphs of sparse graphs (\emph{communities}), which appear in most real-world complex networks, play an important role in many contexts. Computing them however is generally expensive. We propose here a measure of similarities between vertices based on random walks which has 
several important advantages: it captures well the community structure in a network, it can be computed efficiently, and it can be used in an agglomerative algorithm to compute efficiently the community structure of a network. We propose such an algorithm, called {\em Walktrap}, which runs in time $O(mn^2)$ and space $O(n^2)$ in the worst case, and in time $O(n^2\log n)$ and space $O(n^2)$ in most real-world cases ($n$ and $m$ are respectively the number of vertices and edges in the input graph). Extensive comparison tests show that our algorithm surpasses previously proposed ones concerning the quality of the obtained community structures and that it stands among the best ones concerning the running time.}}

\section{Introduction}
Recent advances have brought out the importance of \emph{complex networks} in many different domains such as sociology (acquaintance networks, collaboration networks), biology (metabolic networks, gene networks) or computer science (internet topology, web graph, p2p networks). We refer to \cite{wasserman94socialnetwork,Strogatz:2001,albert:2002,Newman:2003,Dorogovtsev:2003} for reviews from different perspectives and for an extensive bibliography. The associated graphs are in general globally sparse but locally dense: there exist groups of vertices, called \emph{communities}, highly connected between them but with few links to other vertices. This kind of structure brings out much information about the network. For example, in a metabolic network the communities correspond to biological functions of the cell \cite{Ravasz:2002}. In the web graph the communities correspond to topics of interest \cite{kleinberg:2001,Flake:2002}.

This notion of community is however difficult to define formally. Many definitions have been proposed in social networks studies \cite{wasserman94socialnetwork}, but they are too restrictive or cannot be computed efficiently. However, most recent approaches have reached a consensus, and consider that a partition $\mathcal{P} = \{C_1, \dots, C_k\}$ of the vertices of a graph $G = (V,E)$ ($\forall i, C_i \subseteq V$) represents a good community structure if the proportion of edges inside the $C_i$ (internal edges) is high compared to the proportion of edges between them (see for example the definitions given in \cite{Fortunato:2004}). Therefore, we will design an algorithm which finds communities satisfying this criterion. More precisley, we will evaluate the quality of a partition into communities using a quantity (known as {\em modularity} \cite{Newman:2004,Newman_Girvan:2004}) which captures this.

We will consider throughout this paper an \emph{undirected graph} $G = (V,E)$ with $n = |V|$ vertices and $m = |E|$ edges. We impose that each vertex is linked to itself by a loop (we add these loops if necessary). We also suppose that $G$ is connected, the case where it is not being treated by considering the components as different graphs.

\subsection{Our approach and results}

Our approach is based on the following intuition: random walks on a graph tend to get ``trapped'' into densely connected parts corresponding to communities. We therefore begin with some properties of random walks on graphs. Using them, we define a measurement of the structural similarity between vertices and between communities, thus defining a distance. We relate this distance to existing spectral approaches of the problem. But our distance has an important advantage on these methods: it is efficiently computable, and can be used in a hierarchical clustering algorithm (merging iteratively the vertices into communities). One obtains this way a hierarchical community structure that may be represented as a tree called \emph{dendrogram} (an example is provided in Figure~\ref{figure:example}). We propose such an algorithm, called {\em Walktrap}, which computes a community structure in time $\mathcal{O}(mnH)$ where $H$ is the height of the corresponding dendrogram. The worst case is $\mathcal{O}(mn^2)$. But most real-world complex networks are sparse ($m = \mathcal{O}(n)$) and, as already noticed in \cite{Clauset_Newman:2004}, $H$ is generally small and tends to the most favourable case in which the dendrogram is balanced ($H = \mathcal{O}(\log n)$). In this case, the complexity is therefore $\mathcal{O}(n^2\log n)$. We finally evaluate the performance of our algorithm with different experiments which show that it surpasses previously proposed algorithms in most cases.

\subsection{Related work} \label{related_work}

Many algorithms to find community structures in graphs exist. Most of them result from very recent works, but this topic is related to the classical problem of \emph{graph partitioning} that consists in splitting a graph into a given number of groups while minimizing the cost of the edge cut \cite{Fiedler:1973,Pothen:1990,Kernighan:1970}. However, these algorithms are not well suited to our case because they need the number of communities and their size as parameters. The recent interest in the domain has started with a new \emph{divisive} approach proposed by Girvan and Newman \cite{Girvan_Newman:2002, Newman_Girvan:2004}: the edges with the largest \emph{betweenness} (number of shortest paths passing through an edge) are removed one by one in order to split hierarchically the graph into communities. This algorithm runs in time $\mathcal{O}(m^2n)$. Similar algorithms were proposed by Radicchi \textit{et al} \cite{Radicchi_Filippo:2004} and by Fortunato \textit{et al} \cite{Fortunato:2004}. The first one uses a local quantity (the number of loops of a given length containing an edge) to choose the edges to remove and runs in time $\mathcal{O}(m^2)$. The second one uses a more complex notion of information centrality that gives better results but poor performances in $\mathcal{O}(m^3n)$.

\emph{Hierarchical clustering} is another classical approach introduced by sociologists for data analysis \cite{Aldenderfer:1984,Everitt:2001}. From a measurement of the similarity between vertices, an \emph{agglomerative} algorithm groups iteratively the vertices into communities (different methods exist, depending on the way of choosing the communities to merge at each step). Several agglomerative methods have been recently introduced and we will use it in our approach. Newman proposed in \cite{Newman:2004} a greedy algorithm that starts with $n$ communities corresponding to the vertices and merges communities in order to optimize a function called modularity which measures the quality of a partition. This algorithm runs in $\mathcal{O}(mn)$ and has recently been improved to a complexity $\mathcal{O}(mH\log n)$ (with our notations) \cite{Clauset_Newman:2004}. The algorithm of Donetti and Mu\~noz \cite{Donetti:2004} also uses a hierarchical clustering method: they use the eigenvectors of the Laplacian matrix of the graph to measure the similarities between vertices. The complexity is determined by the computation of all the eigenvectors, in $\mathcal{O}(n^3)$ time for sparse matrices. Other interesting methods have been proposed, see for instance \cite{Wu_Huberman:2003, LdaFCosta:2004, Reichardt_Bornholdt:2004, Bagrow_Bollt:2004, Clauset:2005, Duch_Arenas:2005}.

Random walks themselves have already been used to infer structural properties of networks in some previous works. Gaume \cite{Gaume:2004} used this notion in linguistic context. Fouss {\textit et al} \cite{Fouss} used the Euclidean commute time distance based on the average first-passage time of walkers. Zhou and Lipowsky \cite{ZhouL04} introduced another dissimilarity index based on the same quantity; it has been used in a hierarchical algorithm (called {\em Netwalk}). {\em Markov Cluster Algorithm} \cite{mcl} iterates two matrix operations (one corresponding to random walks) bringing out clusters in the limit state. Unfortunately the three last approaches run in $\mathcal{O}(n^3)$ and cannot manage networks with more than a few thousand vertices. Our approach has the main advantage to be significatively faster while producing very good results.

\section{Preliminaries on random walks}

The graph G is associated to its \emph{adjacency matrix} $A$: $A_{ij} = 1$ if vertices $i$ and $j$ are connected and $A_{ij} = 0$ otherwise. The degree $d(i) = \sum_j A_{ij}$ of vertex $i$ is the number of its neighbors (including itself). As we discussed in the introduction, the graph is assumed to be connected. To simplify the notations, we only consider unweighted graphs in this paper. It is however trivial to extend our results to weighted graphs ($A_{ij} \in \mathbb{R}^+$ instead of  $A_{ij} \in \{0,1\}$), which is an advantage of this approach.\\

Let us consider a discrete \emph{random walk process} (or diffusion process) on the graph $G$ (see \cite{Lovasz_random_walks,book_Aldous} for a complete presentation of the topic). At each time step a walker is on a vertex and moves to a vertex chosen randomly and uniformly among its neighbors. The sequence of visited vertices is a \emph{Markov chain}, the states of which are the vertices of the graph. At each step, the transition probability from vertex $i$ to vertex $j$ is $P_{ij} = \frac{A_{ij}}{d(i)}$. This defines the \emph{transition matrix} $P$ of random walk processes. One can also write $P = D^{-1}A$ where $D$ is the diagonal matrix of the degrees ($\forall i, D_{ii} = d(i)$ and $D_{ij} = 0$ for $i \neq j$).

The process is driven by the powers of the matrix $P$: the probability of going from $i$ to $j$ through a random walk of length $t$ is $(P^t)_{ij}$. In the following, we will denote this probability by $P_{ij}^t$. It satisfies two well known properties of the random walk process which we will use in the sequel:

\begin{property} \label{limit_proba}
When the length $t$ of a random walk starting at vertex $i$ tends towards infinity, the probability of being on a vertex $j$ only depends on the degree of vertex $j$ (and not on the starting vertex $i$):
$$\forall i, \lim_{t \rightarrow +\infty} P_{ij}^t = \frac{d(j)}{\sum_{k}d(k)}$$
\end{property}
We will provide a proof of this property in the next section.
\begin{property} \label{symetry_proba}
The probabilities of going from $i$ to $j$ and from $j$ to $i$ through a random walk of a fixed length $t$ have a ratio that only depends on the degrees $d(i)$ and $d(j)$:
$$
\forall i, \forall j, d(i) P_{ij}^t = d(j) P_{ji}^t
$$
\end{property}
\begin{proof}
This property can be written as the matricial equation $DP^tD^{-1} = (P^t)^T$ (where $M^T$ is the transpose of the matrix $M$). By using $P = D^{-1}A$ and the symmetry of the matrices $D$ and $A$, we have: $DP^tD^{-1} = D(D^{-1}A)^tD^{-1} = (AD^{-1})^t = (A^T(D^{-1})^T)^t = ((D^{-1}A)^T)^t = (P^t)^T$.
\end{proof}

\section{Comparing vertices using short random walks}

\subsection{A distance $r$ to measure vertex similarities}

In order to group the vertices into communities, we will now introduce a distance $r$ between the vertices that captures the community structure of the graph. This distance must be large if the two vertices are in different communities, and on the contrary if they are in the same community it must be small. It will be computed from the information given by random walks in the graph.

Let us consider random walks on $G$ of a given length $t$. We will use the information given by all the probabilities $P_{ij}^t$ to go from $i$ to $j$ in $t$ steps. The length $t$ of the random walks must be sufficiently long to gather enough information about the topology of the graph. However $t$ must not be too long, to avoid the effect predicted by Property~\ref{limit_proba}; the probabilities would only depend on the degree of the vertices. Each probability $P_{ij}^t$ gives some information about the two vertices $i$ and $j$, but Property~\ref{symetry_proba} says that $P_{ij}^t$ and $P_{ji}^t$ encode exactly the same information. Finally, the information about vertex $i$ encoded in $P^t$ resides in the $n$ probabilities $(P_{ik}^t)_{1 \leq k \leq n}$, which is nothing but the $i^{\textrm{\tiny{th}}}$ row of the matrix $P^t$, denoted by $P_{i\point}^t$. To compare two vertices $i$ and $j$ using these data, we must notice that:
\begin{itemize}
\item If two vertices $i$ and $j$ are in the same community, the probability $P_{ij}^t$ will surely be high. But the fact that $P_{ij}^t$ is high does not necessarily imply that $i$ and $j$ are in the same community.
\item The probability $P_{ij}^t$ is influenced by the degree $d(j)$ because the walker has higher probability to go to high degree vertices.
\item Two vertices of a same community tend to ``see'' all the other vertices in the same way. Thus if $i$ and $j$ are in the same community, we will probably have $\forall k, P_{ik}^t \simeq P_{jk}^t$.\\
\end{itemize}
We can now give the definition of our distance between vertices, which takes into account all previous remarks:
\begin{definition} Let $i$ and $j$ be two vertices in the graph and 
\begin{equation} \label{definition_distance}
r_{ij} = \sqrt{\sum_{k = 1}^n\frac{(P_{ik}^t - P_{jk}^t)^2}{d(k)}} = \Big\|D^{-\frac{1}{2}}P_{i\point}^t - D^{-\frac{1}{2}}P_{j\point}^t \Big\| 
\end{equation}
where $\|.\|$ is the Euclidean norm of  $\mathbb{R}^n$.
\end{definition}
One can notice that this distance can also be seen as the $L^2$ distance \cite{book_Aldous} between the two probability distributions $P_{i\point}^t$ and $P_{j\point}^t$. Notice also that the distance depends on $t$ and should be denoted by $r_{ij}(t)$. We will however consider it as implicit to simplify the notations.

Now we generalize our distance between vertices to a distance between communities in a straightforward way. Let us consider random walks that start from a community: the starting vertex is chosen randomly and uniformly among the vertices of the community. We define the probability $P_{Cj}^t$ to go from community $C$ to vertex $j$ in $t$ steps:
$$
P_{Cj}^t = \frac{1}{|C|}\sum_{i \in C} P_{ij}^t
$$
This defines a probability vector $P_{C\point}^t$ that allows us to generalize our distance:
\begin{definition}
Let $C_1, C_2 \subset V$ be two communities. We define the distance $r_{C_1C_2}$ between these two communities by:
$$
r_{C_1C_2} = \Big\| D^{-\frac{1}{2}}P_{C_1\point}^t - D^{-\frac{1}{2}}P_{C_2\point}^t \Big\| = \sqrt{\sum_{k = 1}^n\frac{(P_{C_1k}^t - P_{C_2k}^t)^2}{d(k)}}
$$
\end{definition}
This definition is consistent with the previous one: $r_{ij} = r_{\{i\}\{j\}}$ and we can also define the distance between a vertex $i$ and a community $C$: $r_{iC} = r_{\{i\}C}$.
\subsection{Relation with spectral approaches}

\begin{theorem} \label{th_distance}
The distance $r$ is related to the spectral properties of the matrix $P$ by:
$$
r_{ij}^2 = \sum_{\alpha = 2}^n \lambda_{\alpha}^{2t} (v_{\alpha}(i) - v_{\alpha}(j))^2
$$
where $(\lambda_{\alpha})_{1 \leq \alpha \leq n}$ and $(v_{\alpha})_{1 \leq \alpha \leq n}$ are respectively the eigenvalues and right eigenvectors of the matrix $P$.
\end{theorem}
In order to prove this theorem, we need the following technical lemma:
\begin{lemma} \label{lemma}
The eigenvalues of the matrix $P$ are real and satisfy:
$$
1 = \lambda_1 > \lambda_2 \geq \dots \geq \lambda_n > -1
$$
Moreover, there exists an orthonormal family of vectors $(s_{\alpha})_{1 \leq \alpha \leq n}$ such that each vector $v_{\alpha} = D^{-\frac{1}{2}} s_{\alpha}$ and $u_{\alpha} = D^{\frac{1}{2}} s_{\alpha}$ are respectively a right and a left eigenvector associated to the eigenvalue $\lambda_{\alpha}$:
$$
\forall \alpha, P v_{\alpha} = \lambda_{\alpha} v_{\alpha} \textrm{ and } P^T u_{\alpha} = \lambda_{\alpha} u_{\alpha}
$$
$$
\forall \alpha, \forall \beta, v_{\alpha}^T u_{\beta} = \delta_{\alpha\beta}
$$
\end{lemma}
\begin{proof}The matrix $P$ has the same eigenvalues as its similar matrix $S = D^{\frac{1}{2}}PD^{-\frac{1}{2}} = D^{-\frac{1}{2}}AD^{-\frac{1}{2}}$. The matrix $S$ is real and symmetric, so its eigenvalues $\lambda_{\alpha}$ are real. $P$ is a stochastic matrix ($\sum_{j=1}^n P_{ij} = 1$), so its largest eigenvalue is $\lambda_1 = 1$. The graph $G$ is connected and primitive (the $\gcd$ of the cycle lengths of $G$ is $1$, due to the loops on each vertex), therefore we can apply the Perron-Frobenius theorem which implies that $P$ has a unique dominant eigenvalue. Therefore we have: $|\lambda_{\alpha}| < 1$ for $2 \leq \alpha \leq n$.

The symmetry of $S$ implies that there also exists an orthonornal family $s_{\alpha}$ of eigenvectors of $S$ satisfying $\forall \alpha, \forall \beta, s_{\alpha}^Ts_{\beta} = \delta_{\alpha\beta}$ (where $\delta_{\alpha\beta} = 1$ if $\alpha = \beta$ and $0$ otherwise). We then directly obtain that the vectors $v_{\alpha} = D^{-\frac{1}{2}} s_{\alpha}$ and $u_{\alpha} = D^{\frac{1}{2}} s_{\alpha}$ are respectively a right and a left eigenvector of $P$ satisfying $u_{\alpha}^Tv_{\beta} = \delta_{\alpha\beta}$.
\end{proof}\\
We can now prove Theorem~\ref{th_distance} and obtain Property~\ref{limit_proba} as a corrolary:\\

\begin{proof}Lemma \ref{lemma} makes it possible to write a spectral decomposition of the matrix $P$: 
$$
P = \sum_{\alpha = 1}^n \lambda_{\alpha}v_{\alpha}u_{\alpha}^T \textrm{, and } P^t = \sum_{\alpha = 1}^n \lambda_{\alpha}^tv_{\alpha}u_{\alpha}^T \textrm{, and so } P_{ij}^t = \sum_{\alpha = 1}^n \lambda_{\alpha}^tv_{\alpha}(i)u_{\alpha}(j)
$$
When $t$ tends towards infinity, all the terms $\alpha \geq 2$ vanish. It is easy to show that the first right eigenvector $v_1$ is constant. By normalizing we have $\forall i, v_1(i) = \frac{1}{\sqrt{\sum_k d(k)}}$ and $\forall j, u_1(j) = \frac{d(j)}{\sqrt{\sum_k d(k)}}$. We obtain Property~\ref{limit_proba}:
$$
\lim_{t \rightarrow +\infty} P_{ij}^t = \lim_{t \rightarrow +\infty} \sum_{\alpha = 1}^n \lambda_{\alpha}^t v_{\alpha}(i) u_{\alpha}(j) = v_1(i) u_1(j) = \frac{d(j)}{\sum_{k=1}^nd(k)}
$$
Now we obtain the expression of the probability vector $P_{i\point}^t$:
$$
P_{i\point}^t = \sum_{\alpha = 1}^n \lambda_{\alpha}^t v_{\alpha}(i) u_{\alpha} = D^{\frac{1}{2}} \sum_{\alpha = 1}^n \lambda_{\alpha}^t v_{\alpha}(i) s_{\alpha}
$$
We put this formula into the second definition of $r_{ij}$ given in Equation (\ref{definition_distance}). Then we use the Pythagorean theorem with the orthonormal family of vectors $(s_{\alpha})_{1 \leq \alpha \leq n}$, and we remember that the vector $v_1$ is constant to remove the case $\alpha = 1$ in the sum. Finally we have:
$$
r_{ij}^2 = \bigg\|\sum_{\alpha = 1}^n \lambda_{\alpha}^t (v_{\alpha}(i) - v_{\alpha}(j)) s_{\alpha} \bigg\|^2 = \sum_{\alpha = 2}^n \lambda_{\alpha}^{2t} (v_{\alpha}(i) - v_{\alpha}(j))^2
$$
\end{proof}

This theorem relates random walks on graphs to the many current works that study community structure using spectral properties of graphs. For example, \cite{Simonsen:2004} notices that the modular structure of a graph is expressed in the eigenvectors of $P$ (other than $v_1$) that corresponds to the largest positive eigenvalues. If two vertices $i$ and $j$ belong to a same community then the coordinates $v_{\alpha}(i)$ and $v_{\alpha}(j)$ are similar in all these eigenvectors. Moreover, \cite{Schulman:2001,Gaveau:1999} show in a more general case that when an eigenvalue $\lambda_{\alpha}$ tends to $1$, the coordinates of the associated eigenvector $v_{\alpha}$ are constant in the subsets of vertices that correspond to communities. A distance similar to ours (but that cannot be computed directly with random walks) is also introduced: $d_{t}^2(i,j) = \sum_{\alpha = 2}^{n} \frac{(v_{\alpha}(i) - v_{\alpha}(j))^2}{1 - |\lambda_{\alpha}|^t}$. Finally, \cite{Donetti:2004} uses the same spectral approach applied to the Laplacian matrix of the graph $L = D - A$.

All these studies show that the spectral approach takes an important part in the search for community structure in graphs. However all these approaches have the same drawback: the eigenvectors need to be explicitly computed (in time $\mathcal{O}(n^3)$ for a sparse matrix). This computation rapidly becomes untractable in practice when the size of the graph exceeds some thousands of vertices. Our approach is based on the same foundation but has the advantage of avoiding the expensive computation of the eigenvectors: it only needs to compute the probabilities $P_{ij}^t$, which can be done efficiently as shown in the following subsection.

\subsection{Computation of the distance $r$}

Once the two vectors $P_{i\point}^t$ and $P_{j\point}^t$ are computed, the distance $r_{ij}$ can be computed in time $\mathcal{O}(n)$ using Equation (\ref{definition_distance}). Notice that given the probability vectors $P_{C_1\point}^t$ and $P_{C_2\point}^t$, the distance $r_{C_1C_2}$ is also computed in time $\mathcal{O}(n)$

The probability vectors can be computed once and stored in memory (which uses $\mathcal{O}(n^2)$ memory space) or they can be dynamically computed (which increases the time complexity) depending on the amont of available memory. We propose an exact method and an approximated method to compute them.

\paragraph{Exact computation}
\begin{theorem}
Each probability vector $P_{i\point}^t$ can be computed in time $\mathcal{O}(tm)$ and space $\mathcal{O}(n)$.
\end{theorem}
\begin{proof}To compute the vector $P_{i\point}^t$, we multiply $t$ times the vector $P_{i\point}^0$ ($\forall k, P_{i\point}^0(k) = \delta_{ik}$) by the matrix $P$. This direct method is advantageous in our case because the matrix $P$ is generally sparse (for real-world complex networks) therefore each product is processed in time $\mathcal{O}(m)$. The initialization of $P_{i\point}^0$ is done in $\mathcal{O}(n)$ and thus each of the $n$ vectors $P_{i\point}^t$ is computed in time $\mathcal{O}(n + tm) = \mathcal{O}(tm)$. 
\end{proof}
\paragraph{Approximated computation}
\begin{theorem}
Each probability vector $P_{i\point}^t$ can be approximated in time $\mathcal{O}(Kt)$ and space $\mathcal{O}(K)$ with an relative error $\mathcal{O}(\frac{1}{\sqrt{K}})$.
\end{theorem}
\begin{proof}
We compute $K$ random walks of length $t$ starting from vertex $i$. Then we approximate each probability $P_{ik}^t$ by $\frac{N_{ik}}{K}$ where $N_{ik}$ is the number of walkers that ended on vertex $k$ during the $K$ random walks. The Central Limit Theorem implies that this quantity tends toward $P_{ik}^t$ with a speed $\mathcal{O}(\frac{1}{\sqrt{K}})$ when $K$ tends toward infinity. Each random walk computation is done in time $\mathcal{O}(t)$ and constant space hence the overall computation is done in time $\mathcal{O}(Kt)$ and space $\mathcal{O}(K)$.
\end{proof}

The approximated method	is only interresting for very large graphs. In the following we will consider the exact method for the complexity and the experimental evaluation.

\subsection{Generalizing the distance}

We saw that our distance is directly related to the spectral properties of the transition matrix $P$. We show in this section how one can generalize easily and efficiently this distance to use another weighting of the eigenvectors. To achieve this, we only need to define different vectors $\widehat{P}_{i\point}$, all the rest of the approach follows.

\begin{theorem}
Let us consider the generalized distance $\widehat{r}^2_{ij} = \displaystyle\sum_{\alpha = 2}^n f^2(\lambda_{\alpha})  (v_{\alpha}(i) - v_{\alpha}(j))^2$ where $f(x) = \displaystyle\sum_{k=0}^{\infty} c_k x^k$ is any function defined by a power series.\\
Then $\widehat{r}_{ij} = \Big\|D^{-\frac{1}{2}}\widehat{P}_{i\point} - D^{-\frac{1}{2}}\widehat{P}_{j\point} \Big\|$, where $\widehat{P}_{i\point} = \sum_{k=0}^{\infty} c_k P_{i\point}^k$, can be approximated in time $\mathcal{O}(rm)$ and space $\mathcal{O}(n)$ with relative error on each coordinate less than $\varepsilon_r = \displaystyle\sum_{k=r+1}^{\infty} c_k$.
\end{theorem}

\begin{proof}
We have $\widehat{P}_{i\point} = \sum_{k=0}^{\infty} c_k P_{i\point}^k = D^{\frac{1}{2}}\sum_{k=0}^{\infty} \sum_{\alpha = 1}^n c_k \lambda_{\alpha}^k v_{\alpha}(i) s_{\alpha}$. Therefore :
$$\widehat{r}_{ij} = \Big\|D^{-\frac{1}{2}}\widehat{P}_{i\point} - D^{-\frac{1}{2}}\widehat{P}_{j\point} \Big\| = \Big\| \sum_{k=0}^{\infty} \sum_{\alpha = 2}^n c_k \lambda_{\alpha}^k (v_{\alpha}(i) - v_{\alpha}(j)) s_{\alpha} \Big\|$$
And we can conclude because the vectors $s_{\alpha}$ are orthonormal :
$$\sum_{\alpha = 2}^n \Big\| \sum_{k=0}^{\infty} c_k \lambda_{\alpha}^k (v_{\alpha}(i) - v_{\alpha}(j)) s_{\alpha} \Big\|^2 = \sum_{\alpha = 2}^n f^2(\lambda_{\alpha}) (v_{\alpha}(i) - v_{\alpha}(j))^2 = \widehat{r}_{ij}^2$$
To compute the vectors, we approximate the series to the order $r$: $\widehat{P}_{i\point} \simeq \sum_{k=0}^{r} c_k P_{i\point}^k$. We only need to compute the successive powers $\widehat{P}_{i\point}^k$ for $0 \leq k \leq r$ which can be done in time $\mathcal{O}(rm)$ and space $\mathcal{O}(n)$.
\end{proof}

To illustrate this generalization, we show that it directly allows to consider continuous random walks. Indeed, the choice of the length of the random walks (which must be an integer) may be restrictive in some cases. To overcome this constraint, one may consider the continuous random walk process: during a period $dt$ the walker will go from $i$ to $j$ with probability $P_{ij}dt$. One can prove that the probabilities to go from $i$ to $j$ after a time $t$ are given by the matrix $e^{t(P - Id)}$. For a given period length $t$, the associated distance is now $\widehat{r}^2_{ij} = \sum_{\alpha = 2}^n e^{2t(\lambda_{\alpha}-1)} (v_{\alpha}(i) - v_{\alpha}(j))^2$ which corresponds to a function $f(x) = e^{t(x-1)} = \sum_{k=0}^{\infty} c_k x^k$ with $c_k = \frac{t^ke^{-t}}{k!}$.

\section{The algorithm}

In the previous section, we have proposed a distance between vertices (and between sets of vertices) to capture structural similarities between them. The problem of finding communities is now a clustering problem. We will use here an efficient hierarchical clustering algorithm that allows us to find community structures at different scales. We present an agglomerative approach based on Ward's method \cite{Ward:1963} that is well suited to our distance and gives very good results while reducing the number of distance computations.

\begin{figure}[!h]
\begin{center}
\includegraphics{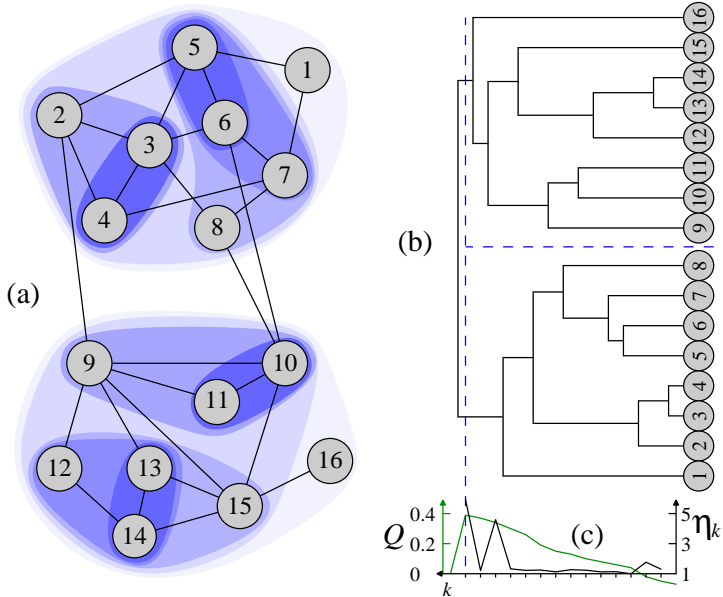}
\caption{(a) An example of community structure found by our algorithm using random walks of length $t = 3$. (b) The stages of the algorithm encoded as a tree called \emph{dendrogram}. The maximum of $\eta_k$ and $Q$, plotted in (c), show that the best partition consists in two communities. The maximal values of $\eta_k$ show also that communities of different scales may be relevant.}
\label{figure:example}
\end{center}
\end{figure}

We start from a partition $\mathcal{P}_1 = \{\{v\}, v \in V\}$ of the graph into $n$ communities reduced to a single vertex. We first compute the distances between all adjacent vertices. Then this partition evolves by repeating the following operations. At each step $k$:
\begin{itemize}
\item choose two communities $C_1$ and $C_2$ in $\mathcal{P}_k$ according to a criterion based on the distance between the communities that we detail later,
\item merge these two communities into a new community $C_3 = C_1 \cup C_2$ and create the new partition: $\mathcal{P}_{k+1} = (\mathcal{P}_k \setminus \{C_1, C_2\}) \cup \{C_3\}$, and
\item update the distances between communities (we will see later that we actually only do this for {\em adjacent} communities).
\end{itemize}
After $n-1$ steps, the algorithm finishes and we obtain $\mathcal{P}_n = \{V\}$. Each step defines a partition $\mathcal{P}_k$ of the graph into communities, which gives a hierarchical structure of communities called dendrogram (see Figure~\ref{figure:example}(b)). This structure is a tree in which the leaves correspond to the vertices and each internal node is associated to a merging of communities in the algorithm: it corresponds to a community composed of the union of the communities corresponding to its children. 

The key points in this algorithm are the way we choose the communities to merge, and the fact that the distances can be updated efficiently. We will also need to evaluate the quality of a partition in order to choose one of the $\mathcal{P}_k$ as the result of our algorithm. We will detail these points below, and explain how they can be managed to give an efficient algorithm.

\subsection{Choosing the communities to merge.} This choice plays a central role for the quality of the obtained community structure. In order to reduce the complexity, we will only merge {\em adjacent} communities (having at least an edge between them). This reasonable heuristic (already used in \cite{Newman:2004} and \cite{Donetti:2004}) limits to $m$ the number of possible mergings at each stage. Moreover it ensures that each community is connected.

We choose the two communities to merge according to Ward's method. At each step $k$, we merge the two communities that minimize the mean $\sigma_k$ of the squared distances between each vertex and its community. 
$$
\sigma_k = \frac{1}{n} \sum_{C \in \mathcal{P}_k} \sum_{i\in C} r_{iC}^2
$$
This approach is a greedy algorithm that tries to solve the problem of maximizing $\sigma_k$ for each $k$. This problem is known to be NP-hard: even for a given $k$, maximizing $\sigma_k$ is the NP-hard ``K-Median clustering problem'' \cite{Fernandez_de_la_Vega:2003, Drineas:2004} for $K = (n-k)$ clusters. The existing approximation algorithms \cite{Fernandez_de_la_Vega:2003, Drineas:2004} are exponential with the number of clusters to find and unsuitable for our purpose. 
So for each pair of adjacent communities $\{C_1,C_2\}$, we compute the variation $\Delta\sigma(C_1,C_2)$ of $\sigma$ that would be induced if we merge $C_1$ and $C_2$ into a new community $C_3 = C_1 \cup C_2$. This quantity only depends on the vertices of $C_1$ and $C_2$, and not on the other communities or on the step $k$ of the algorithm:
\begin{equation} \label{def_delta_sigma}
\Delta\sigma(C_1,C_2) = \frac{1}{n}\Big(\sum_{i\in C_3} r_{iC_3}^2 - \sum_{i\in C_1} r_{iC_1}^2 - \sum_{i\in C_2} r_{iC_2}^2\Big)
\end{equation}
Finally, we merge the two communities that give the lowest value of $\Delta\sigma$. 

\subsection{Computing $\Delta\sigma$ and updating the distances.} The important point here is to notice that these quantities can be efficiently computed thanks to the fact that our distance is a Euclidean distance, which makes it possible to obtain the two following classical results \cite{Jambu}:
\begin{theorem} \label{th_delta_sigma_1} The increase of $\sigma$ after the merging of two communities $C_1$ and $C_2$ is directly related to the distance $r_{C_1C_2}$ by:
$$
\Delta\sigma(C_1,C_2) = \frac{1}{n}\frac{|C_1||C_2|}{|C_1| + |C_2|} r_{C_1C_2}^2
$$
\end{theorem}
\begin{proof}
First notice that $\sum_{i\in C_1} (P_{C_1\point}^t - P_{i\point}^t) = 0$ and $(|C_1| + |C_2|) P_{C_3\point}^t = |C_1| P_{C_1\point}^t + |C_2| P_{C_2\point}^t$. Then we consider the distance $r$ as a metric in $\mathbb{R}^n$ (that contains the probability vectors $P_{C\point}$) associated to an inner product $<.|.>$. Finally, after some elementary computations, we obtain :
$$
\sum_{i\in C_1} r_{iC_3}^2 =  \sum_{i\in C_1} <P_{C_3\point}^t - P_{i\point}^t|P_{C_3\point}^t - P_{i\point}^t> = \sum_{i\in C_1} r_{iC_1}^2 + \frac{|C_1||C_2|^2}{(|C_1| + |C_2|)^2} r_{C_1C_2}^2
$$
This also holds if we replace $C_1$ by $C_2$ and $C_2$ by $C_1$. Therefore:
$$
\sum_{i\in C_3} r_{iC_3}^2 = \sum_{i\in C_1} r_{iC_3}^2 + \sum_{i\in C_2} r_{iC_3}^2 = \sum_{i\in C_1} r_{iC_1}^2 + \sum_{i\in C_2} r_{iC_2}^2 + \frac{|C_1||C_2|}{|C_1| + |C_2|} r_{C_1C_2}^2
$$
We deduce the claim by replacing this expression into Equation (\ref{def_delta_sigma}).
\end{proof}

This theorem shows that we only need to update the distances between communities to get the values of $\Delta\sigma$: if we know the two vectors $P_{C_1\point}$ and $P_{C_2\point}$, the computation of $\Delta\sigma(C_1,C_2)$ is possible in $\mathcal{O}(n)$. Moreover, the next theorem shows that if we already know the three values $\Delta\sigma(C_1,C_2)$, $\Delta\sigma(C_1,C)$ and $\Delta\sigma(C_2,C)$, then we can compute $\Delta\sigma(C_1 \cup C_2, C)$ in constant time.

\begin{theorem}[Lance-Williams-Jambu formula] \label{th_delta_sigma_2} If $C_1$ and $C_2$ are merged into $C_3 = C_1 \cup C_2$ then for any other community $C$:
\begin{equation} \label{eq_th_delta_sigma_2}
\Delta\sigma(C_3, C) = \frac{(|C_1| + |C|) \Delta\sigma(C_1,C) + (|C_2| + |C|) \Delta\sigma(C_2,C) - |C| \Delta\sigma(C_1,C_2)}{|C_1|+|C_2|+|C|}
\end{equation}
\end{theorem}
\begin{proof}
We replace the four $\Delta\sigma$ of Equation~(\ref{eq_th_delta_sigma_2}) by their values given by Theorem~\ref{th_delta_sigma_1}. We multiply each side by $\frac{n(|C_1|+|C_2|+|C|)}{|C|}$ and use $|C_3| = |C_1|+|C_2|$, and obtain the equivalent equation:
$$
(|C_1|+|C_2|)r_{C_3C}^2 = |C_1|r_{C_1C}^2 + |C_2|r_{C_2C}^2 - \frac{|C_1||C_2|}{|C_1|+|C_2|}r_{C_1C_2}^2
$$
Then we use the fact that $P_{C_3\point}^t$ is the barycenter of $P_{C_1\point}^t$ weighted by $|C_1|$ and of $P_{C_2\point}^t$ weighted by $|C_2|$, therefore: 
$$
|C_1|r_{C_1C}^2 + |C_2|r_{C_2C}^2 = (|C_1|+|C_2|)r_{C_3C}^2 + |C_1|r_{C_1C_3}^2 + |C_2|r_{C_2C_3}^2
$$
We conclude using $|C_1|r_{C_1C_3}^2 + |C_2|r_{C_2C_3}^2 = \frac{|C_1||C_2|}{|C_1|+|C_2|}r_{C_1C_2}^2$.
\end{proof}

Since we only merge adjacent communities, we only need to update the values of $\Delta\sigma$ between adjacent communities (there are at most $m$ values). These values are stored in a balanced tree in which we can add, remove or get the minimum in $\mathcal{O}(\log m)$. Each computation of a value of $\Delta\sigma$ can be done in time $\mathcal{O}(n)$ with Theorem~\ref{th_delta_sigma_1} or in constant time when Theorem~\ref{th_delta_sigma_2} can be applied.

\subsection{Evaluating the quality of a partition.} The algorithm induces a sequence $(\mathcal{P}_k)_{1 \leq k \leq n}$ of partitions into communities. We now want to know which partitions in this sequence capture well the community structure. The most widely used criterion is the modularity $Q$ introduced in \cite{Newman:2004,Newman_Girvan:2004}, which relies on the fraction of edges $e_C$ inside community $C$ and the fraction of edges\footnote{inter-community edges contribute for $\frac{1}{2}$ to each community.} $a_C$ bound to community~$C$:
$$
Q(\mathcal{P}) = \sum_{C \in \mathcal{P}} e_C - a_C^2
$$
The best partition is then considered to be the one that maximizes $Q$.

However, depending on one's objectives, one may consider other quality criterion of a partition into communities. For instance, the modularity is not well suited to find communities at different scales. Here we provide another criterion that helps in finding such structures. When we merge two very different communities (with respect to the distance $r$), the value $\Delta\sigma_k = \sigma_{k+1} - \sigma_{k}$ at this step is large. Conversely, if $\Delta\sigma_k$ is large then the communities at step $k-1$ are surely relevant. To detect this, we introduce the increase ratio $\eta_k$:
$$
\eta_k = \frac{\Delta\sigma_k}{\Delta\sigma_{k-1}} = \frac{\sigma_{k+1} - \sigma_k}{\sigma_k - \sigma_{k-1}}
$$
One may then consider that the relevant partitions $\mathcal{P}_k$ are those associated with the largest values of $\eta_k$. Depending on the context in which our algorithm is used, one may take only the best partition (the one for which $\eta_k$ is maximal) or choose among the best ones using another criterion (like the size of the communities, for instance). This is an important advantage of our method, which helps in finding the different scales in the community structure. However we used the modularity (which produces better results to find an unique partition and is not specific to our algorithm) in our experimental tests to be able to compare our algorithm with the previouly proposed ones.

\subsection{Complexity.} 
First, the initialization of the probability vectors is done in $\mathcal{O}(mnt)$. Then, at each step $k$ of the algorithm, we keep in memory the vectors $P_{C\point}^t$ corresponding to the current communities (the ones in the current partition). But for the communities that are not in $\mathcal{P}_k$  (because they have been merged with another community before) we only keep the information saying in which community it has been merged. We keep enough information to construct the dendogram and have access to the composition of any community with a few more computation.

\noindent
When we merge two communities $C_1$ and $C_2$ we perform the following operations:
\begin{itemize}
\item Compute $P_{(C_1\cup C_2)\point}^t = \frac{ |C_1| P_{C_1\point}^t +  |C_2| P_{C_2\point}^t}{ |C_1|+|C_2|}$ and remove $ P_{C_1\point}^t$ and $P_{C_2\point}^t$.
\item Update the values of $\Delta\sigma$ concerning $C_1$ and $C_2$ using Theorem~\ref{th_delta_sigma_2} if possible, or otherwise using Theorem~\ref{th_delta_sigma_1}.
\end{itemize}
The first operation can be done in $\mathcal{O}(n)$, and therefore does not play a significant role in the overall complexity of the algorithm. The dominating factor in the complexity of the algorithm is the number of distances $r$ computed (each one in $\mathcal{O}(n)$). We prove an upper bound of this number that depends on the height of the dendrogram. We denote by $h(C)$ the height of a community $C$ and by $H$ the height of the whole tree ($H = h(V)$).

\begin{theorem}
An upper bound of the number of distances computed by our algorithm is $2mH$. Therefore its global time complexity is $\mathcal{O}(mn(H+t))$.
\end{theorem}
\begin{proof}Let $M$ be the number of computations of $\Delta\sigma$. $M$ is equal to $m$ (initialization of the first $\Delta\sigma$) plus the sum over all steps $k$ of the number of neighbors of the new community created at step $k$ (when we merge two communities, we need to update one value of $\Delta\sigma$ per neighbor). For each height $1 \leq h \leq H$, the communities with the same height $h$ are pairwise disjoint, and the sum of their number of neighbor communities is less than $2m$ (each edge can at most define two neighborhood relations). The sum over all heights finally gives $M \leq 2Hm$. Each of these $M$ computations needs at most one computation of $r$ in time $\mathcal{O}(n)$ (Theorem~\ref{th_delta_sigma_1}). Therefore, with the initialization, the global complexity is $\mathcal{O}(mn(H+t))$.
\end{proof}

In practice, a small $t$ must be chosen (we must have $t = \mathcal{O}(\log n)$ due to the exponential convergence speed of the random walk process) and thus the global complexity is $\mathcal{O}(mnH)$. We always empirically observed that best results are obtained using length $3 \leq t \leq 8$. We moreover observed that the choice of $t$ in this range is not crutial as the results are often similar. Hence we think that a good empirical compromise is to choose $t = 4$ or $t = 5$. We also advise to reduce this length for very dense graphs and to increase it for very sparse ones because the convergence speed of the random walk process increase with the graph density. Studying more formally the influence of $t$, and determining optimal values, remains to be done.

 The worst case is $H = n-1$, which occurs when the vertices are merged one by one to a large community. This happens in the ``star'' graph, where a central vertex is linked to the $n-1$ others. However Ward's algorithm is known to produce small communities of similar sizes. This tends to get closer to the favorable case in which the community structure is a balanced tree and its height is $H = \mathcal{O}(\log n)$. 

However, this upper bound is not reached in practical cases. We evaluated the actual number of distance computations done on graphs from the test set presented in Section~\ref{comparison_generated}. We chose graphs with $n=3\,000$ vertices, their mean number of edges is $m=47\,000$ and the mean height of the computed dendrograms is $H=31.6$. We compared the worst case upper bound $2(mn(n-1))$ and the upper bound $2mnH$ with the actual number distances computed with and without using Theorem~\ref{th_delta_sigma_2}. 

We also considered an additional heuristics that consists in applying Theorem~\ref{th_delta_sigma_2} whenever we only know one of the two quantities $\Delta\sigma(C_1,C)$ or $\Delta\sigma(C_2,C)$. In this case we assume that the other one is greater than the current minimal $\Delta\sigma$ and we obtain a lower bound for $\Delta\sigma(C_1 \cup C_2,C)$. Later, if this lower bound becomes the minimal $\Delta\sigma$ then we compute the exact distance in $\mathcal{O}(n)$. Otherwise if the community $C_3 = C_1 \cup C_2$ is merged using another community than $C$ the exact computation is avoided. This heuristics can induce inexact merging ordering when the other unknown $\Delta\sigma$ is not greater than the current minimal $\Delta\sigma$, we observed in this test that this happened on $0.05\%$ of the cases. 

The results, transcribed in Table~\ref{table:distances}, show that in practical cases, the actual complexity of our approach is significantly lower than the upper bound we proved. However, this upper bound can be reached in the pathological case of the star graph.

\begin{table} 
\begin{center}
\begin{tabular}{|c|c|c|}
\hline
\multicolumn{2}{|c|}{Method}&Number of distances computed\\
\hline
\multirow{2}{*}{Upper bounds}&$2m(n-1)$&282\,000\,000\\
&$2mH$&2\,970\,000\\
\hline
\multirow{3}{*}{Practical tests}&without theorem \ref{th_delta_sigma_2}&321\,000\\
&with theorem \ref{th_delta_sigma_2}&277\,000\\
&with additional heuristics&103\,000\\
\hline
\end{tabular}
\caption{Number of distances computed according to upper bounds and practical tests.}
\label{table:distances}
\end{center}
\end{table}

\section{Experimental evaluation of the algorithm} \label{experimental_results}

In this section we will evaluate and compare the performances of our algorithm with most previously proposed methods. This comparison has been done in both randomly generated graphs with communities and real world networks. In order to obtain rigorous and precise results, all the programs have been extensively tested on the same large set of graphs.

\noindent The test compares the following community detection programs: 
\begin{itemize}
\item this paper (Walktrap) with random walk length $t = 5$ and $t = 2$,
\item the Girvan Newman algorithm \cite{Girvan_Newman:2002, Newman_Girvan:2004} (a divisive algorithm that removes larger betweeness edges),
\item the Fast algorithm that optimize the modularity proposed by Newman and improved in \cite{Clauset_Newman:2004} (a greedy algorithm designed for very large graphs that optimizes the modularity),
\item the approach of Donetti and Mu\~noz using the Laplacian matrix \cite{Donetti:2004} and its new improved version \cite{Donetti:2005} (a spectral approach with a hierarchical algorithm),
\item the Netwalk algorithm \cite{ZhouL04} (another algorithm based on random walks),
\item the Markov Cluster Algorithm (MCL) \cite{mcl} (an algorithm based on simulation of (stochastic) flow in graphs),
\item and the Cosmoweb algorithm \cite{cosmoweb} (a gravitational approach designed for web clustering).
\end{itemize}
We refer to Section \ref{related_work} and to the cited references for more details on these algorithms.
\subsection{Comparison on generated graphs} \label{comparison_generated}

\begin{figure}[!h]
\begin{center}
\includegraphics{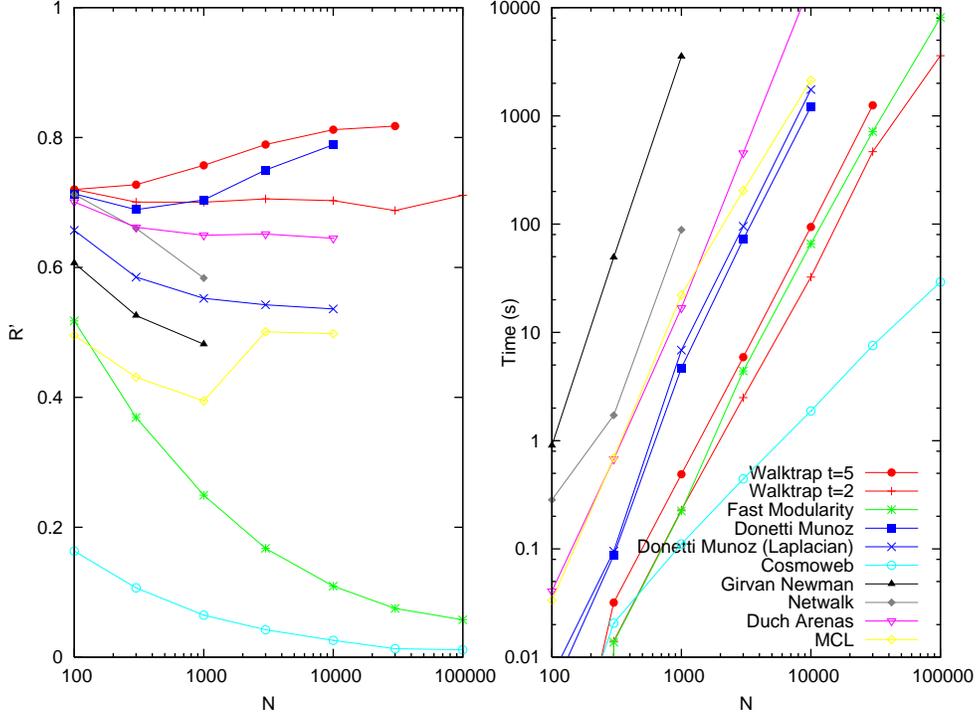}
\caption{Quality and time performance of different approaches in function of the size of the graphs ($N$). (Left) Mean quality of the partition found ($R'$). Right: Mean execution time (in seconds).}
\label{figure:time}
\end{center}
\end{figure}

Evaluating a community detection algorithm is a difficult task because one needs some test graphs whose community structure is already known. A classical approach is to use randomly generated graphs with communities. Here we will use this approach and generate the graphs as follows.

The parameters we consider are :
\begin{itemize}
\item the number $k$ of communities and their sizes $|C_i|$ (these parameters give the number of vertices $N$), 
\item the internal degree $d_{in}(C_i)$ of each community,
\item and the wanted modularity $Q$.
\end{itemize}
In order to reduce the number of parameters, we consider that the external degrees are proportional to the internal degrees: $\forall i,\ d_{out}(C_i) = \beta \times d_{in}(C_i)$. One can check that the expected modularity is then:

$$
Q_e = \frac{1}{1 + \beta} - \frac{\sum_i (d_{in}(C_i)\times|C_i|)^2}{\left(\sum_i d_{in}(C_i)\times|C_i|\right)^2}
$$
We therefore obtain the wanted modularity by choosing the appropriate value for $\beta$.

Once these parameters have been chosen, we draw each internal edge of a given community with the same probability, producing Erd\"os-Renyi like communities. Then the external degrees are chosen proprotionally to the internal degrees (with a factor $\beta$) and the vertices are randomly linked with respect to some constraints (no loop, no multiple edge).

To evaluate the quality of the partition found by the algorithms, we compare them to the original generated partition. To achieve this, we use the Rand index corrected by Hubert and Arabie \cite{Rand, Hubert_Arabie} which evaluates the similarities between two partitions. The Rand index $R(\mathcal{P}_1,\mathcal{P}_2)$ is the ratio of pairs of vertices correlated by the partitions $\mathcal{P}_1$ and $\mathcal{P}_2$ (two vertices are correlated by the partitions $\mathcal{P}_1$ and $\mathcal{P}_2$ if they are classified in the same community or in different communities in the two partitions). The expected value of $R$ for a random partition is not zero. To avoid this, Hubert and Arabie proposed a corrected index that is also more sensitive : $R' = \frac{R - R_{exp}}{R_{max} - R_{exp}}$ where $R_{exp}$ is the expected value of $R$ for two random partitions with the same community size as $\mathcal{P}_1$ and $\mathcal{P}_2$. This quantity can be efficiently computed using the following equivalent formula :
$$
R'(\mathcal{P}_1,\mathcal{P}_2) = \frac{N^2 \displaystyle\sum_{i,j}|C^1_i \cap C^2_j|^2 - \displaystyle\sum_i |C^1_i|^2 \displaystyle\sum_j |C^2_j|^2}{\frac{1}{2}N^2\left(\displaystyle\sum_i |C^1_i|^2 + \displaystyle\sum_j |C^2_j|^2\right) - \displaystyle\sum_i |C^1_i|^2 \displaystyle\sum_j |C^2_j|^2}
$$
Where $(C^x_i)_{1 \leq i \leq k_x}$ are the communities of the partition $\mathcal{P}_x$ and $N$ is the total number of vertices.

This quantity has many advantages compared to the ``ratio of vertices correctly identified'' that has been widely used in the past. It captures the similarities between partitions even if they do not have the same number of communities, which is crucial here as we will see below. Moreover, a random partition always gives the same expected value $0$ that does not depend on the number of communities.

We also compared the partitions using the modularity. However, the results and the conclusions were very similar to those obtained with $R'$. In order to reduce the size of this section and to avoid duplicated information, we only plotted the results obtained with the corrected Rand index $R'$.

\paragraph{Homogeneous graphs}

\begin{figure}[!h]
\begin{center}
\includegraphics{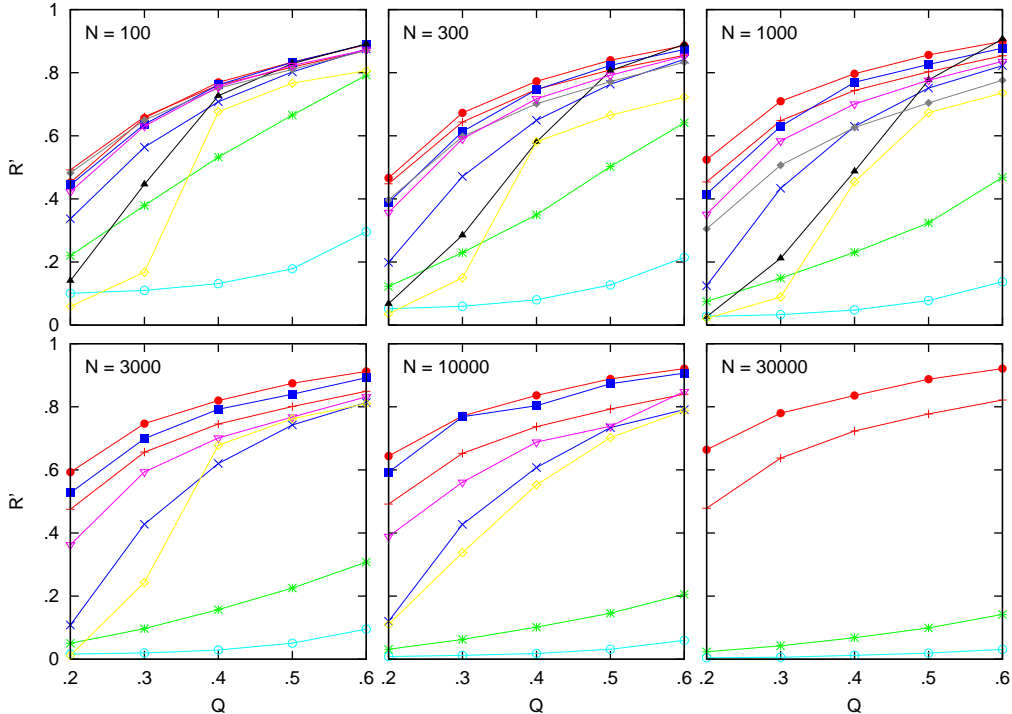}
\caption{Quality of the partition found in function of the modularity of the generated partition for different sizes $N$ (same legend as Figure~\ref{figure:time}).}
\label{figure:size}
\end{center}
\end{figure}

Let us start with the most simple case where all the communities are similar (same size and same density). Therefore we only have to choose the size $N$ of the graphs, the number $k$ of communities, the internal degree $d_{in}$ of communities and the wanted modularity $Q$. The internal edges are drawn with the same probability, producing a Poisson degree distribution. We generated graphs corresponding to combinations of the following parameters:
\begin{itemize}
\item sizes $N$ in $\{100, 300, 1\,000, 3\,000, 10\,000, 30\,000, 100\,000\}$,
\item number of communities, $k = N^{\gamma}$ with $\gamma$ in $\{0.3, 0.42, 0.5\}$,
\item internal degree, $d_{in}(C_i) = \alpha \ln(|C_i|)$ with $\alpha$ in $\{2, 4, 6, 8, 10\}$,
\item wanted modularity $Q$ in $\{0.2, 0.3, 0.4, 0.5, 0.6\}$.
\end{itemize}

The first comparison of the quality and time performances is plotted on Figure~\ref{figure:time}. For each graph size, we plotted the mean corrected Rand index ($R'$) and the mean running time. To avoid that some approaches can be advantaged (or disadvantaged) by particular parameters, the mean has been computed over all the possible combinations of the parameters listed above. This first comparison shows that our algorithm has the advantage of being efficient regarding both the quality of the results and the speed, while other alorithms only achieve one of these goals. It can handle very large graphs with up to 300\,000 vertices (this limitation is due to its memory requirements). Larger graphs can be processed (without the same quality of results) with the Fast Modularity algorithm that has been able to process a 2 million vertex graph.

We also plotted $R'$ on Figure~\ref{figure:size} to observe the influence of the modularity of the generated partition on the results.
These first tests show that most previously proposed approaches have good performances on small graphs. But our approach is the only one that allows to process large graphs while producing good results. Notice that the improved approach of Donetti and Mu\~noz also produces very good results but requires more computational time. This improved version \cite{Donetti:2005} uses exactly the same eigen vectors as the ones we use in our algorithm, which explains that the quality of the results are similar. The MCL algorithm was difficult to use in this intensive test since the user must choose a granularity parameter for each input graph, which is a limitation of this algorithm. We manually chose one parameter for each size of graph (hence the results are not optimal and it can explain their fluctuations), doing our best to find a good one.

\begin{figure}[!h]
\begin{center}
\includegraphics{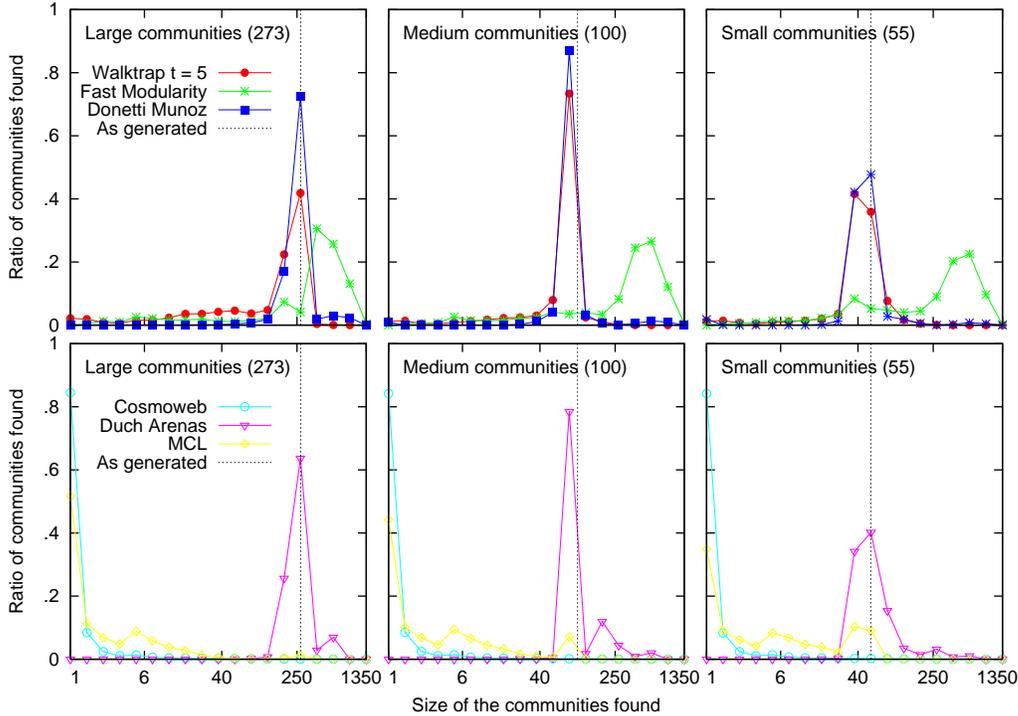}
\caption{Distribution of the size of the communities for three different numbers of generated communities corresponding to 11, 30 or 55 communities on $N = 3000$ vertex graphs.}
\label{figure:com_size}
\end{center}
\end{figure}

It is also interesting to compare the distribution of the size of the communities found to the size of the generated communities. We plotted these quantities on Figure~\ref{figure:com_size} for graphs with $N = 3000$ vertices. We generated graphs with three different sizes of communities and the results can explain the limitations of some approaches. It seems for instance that the Fast Modularity algorithm~\cite{Clauset_Newman:2004} produces communities that always have the same size independantly of the actual size of the communities. Likewise, Cosmoweb~\cite{cosmoweb} produces too many very small communities (1 to 4 vertices). 

\paragraph{Heterogeneous graphs}

\begin{figure}[!h]
\begin{center}
\includegraphics{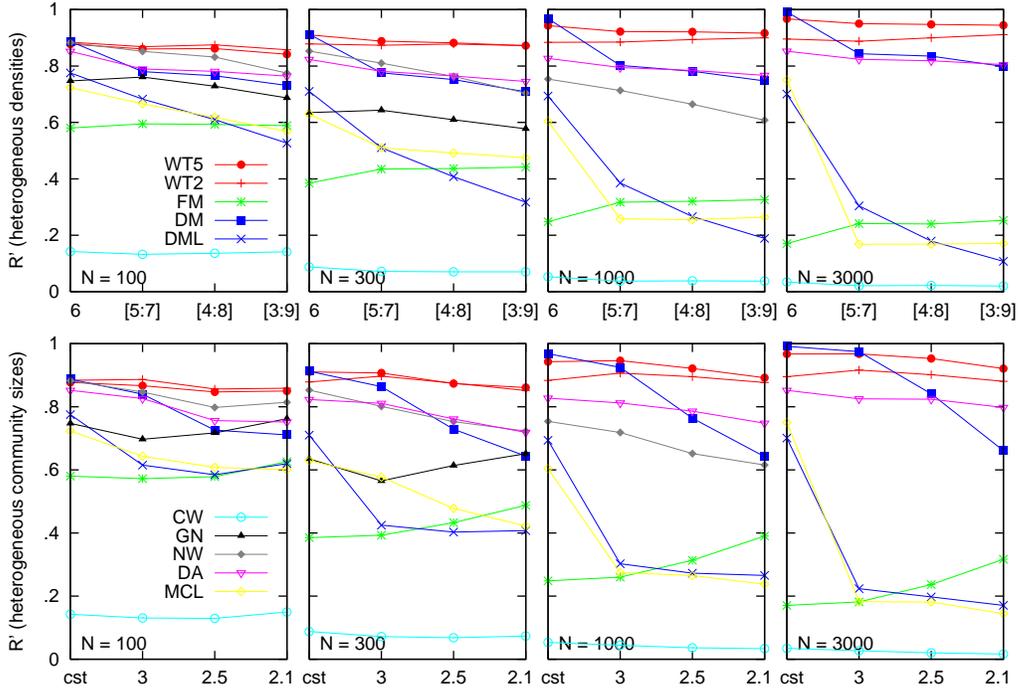}
\caption{Influence of the heterogeneity of the graphs (for four sizes of graphs $N = 100, 300, 1\,000, 3\,000$). On the x axis, left corresponds to homogeneous graphs and right corresponds to very heterogeneous graphs. The quality of the partition is plotted as a function of different parameters as described in the text. Top: internal density given by the range [$\alpha_{min}:\alpha_{max}$]. Bottom: community sizes given by the exponent of the power law distribution.}
\label{figure:heterogeneity}
\end{center}
\end{figure}

The second set of graphs has different kind of communities (different sizes and different densities). The sizes of the communities are randomly chosen according to a power law and the internal densities of each community is also randomly chosen. We therefore have the two following additional parameters:
\begin{itemize}
\item the range of internal degree, $d_{in}(C_i)$ is uniformely chosen between $\alpha_{min} \ln(|C_i|)$ and $\alpha_{max} \ln(|C_i|)$ with $(\alpha_{min}, \alpha_{max}) = (5,7)$, $(4,8)$ and $(3,9)$, and
\item the community size distribution is a power law of exponent $\alpha$ in 2.1, 2.5 and 3.\footnote{The community sizes are chosen within a range $[S_{min}..S_{max}]$ and the probability that a community has size $S$ is actualy proportional to $(S + \mu)^{\alpha}$, with $\mu$ chosen such that the expected size of the overall graph is equal to a given $N$.} 
\end{itemize}
To study the influence of the heterogeneity of the communities, we generated graphs of size $N = 3000$ with all combinations of the previous parameters (modularity, number of communities) and of the two new ones. The three values of the above parameters correspond to three levels of heterogeneity. Figure~\ref{figure:heterogeneity} shows that our approach is not influenced by the heterogeneity of the communities, whereas the others are.

\subsection{Comparison on real world networks}

To extend the comparison between algorithms, we also conducted experiments on some real world networks. However judging the quality of the different partiton found is very difficult because we do not have a reference partition that can be considered as the actual communities of the network. We only compared the value of the modularity found by the different algorithms. The results are reported in Table~\ref{table:reel}. 

\begin{table}[!h]
\begin{center}
\small
\begin{tabular}{|c|c|c|c|c|c|c|}
\hline
graph&karate&foot&protein&arxiv&internet&www\\
\hline
nb vertices/mean degree&33/4.55&115/10.7&594/3.64&9377/5.14&67882/8.12&159683/11.6\\
\hline
\hline
Walktrap (t = 5)&0.38/0s&0.60/0s&0.67/0.02s&0.76/4.61s&0.76/1030s&0.91/5770s\\
\hline
Walktrap (t = 2)&0.38/0s&0.60/0s&0.64/0.01s&0.71/1.08s&0.69/273s&0.84/468s\\
\hline
Fast Modularity&0.39/0s&0.57/0s&0.71/0s&0.77/1.65s&0.72/483s&0.92/1410s\\
\hline
Donetti Mu\~noz&0.41/0s&0.60/0s&0.59/0.34s&0.66/1460s& -- & -- \\
\hline 
Donetti Mu\~noz (Laplacian)&0.41/0s&0.60/0s&0.60/1.37s&0.62/1780s& -- & -- \\
\hline 
Cosmoweb&-0.05/0s&0.33/0s&0.50/0.02s&0.60/0.65&0.47/6.82s&0.79/21s\\
\hline
Girvan Newman&0.40/0s&0.60/0.39s&0.70/6.93s&$>$40000s& -- & -- \\
\hline
Netwalk&0.40/0.02s&0.60/0.07s&0.60/5.2s&$>$40000s& -- & -- \\
\hline
Duch Arenas&0.41/0s&0.60/0.05s&0.69/1.9s&0.77/14000s& -- & -- \\
\hline
MCL&0.36/0s&0.60/0.05s&0.66/0.58s&0.73/61.3s& -- & -- \\
\hline
\end{tabular}
\normalsize
\caption{Performances on real world networks (modularity / time (in seconds)). The second line shows the size of the graphs given by their number of vertices and their mean degree.}
\label{table:reel}
\end{center}
\end{table}

We used the following real world networks :
\begin{itemize}
\item The Zachary's karate club network \cite{Zachary}, a small social network that has been widely used to test most of the community detection algorithms.
\item The college football network from \cite{Girvan_Newman:2002}.
\item The protein interaction network studied in \cite{protein}. 
\item A scientists collabaration network computed on the arXiv database \cite{url_data_arxiv}.
\item An internet map provided by Damien Magoni \cite{Magoni}.
\item The web graph studied in \cite{Albert:1999}
\end{itemize}
We reduced the sizes of these networks by only keeping the largest connected component and by iteratively removing all the one-degree vertices (which do not provide significant information on community structures). This allowed us to run the comparison tests with all the algorithms on smaller networks (Table~\ref{table:reel} reports the size and the mean degree of the graphs after this processing).

\section{Conclusion and further work}

We proposed a new distance between vertices that quantify their structural similarity using random walks. This distance has several advantages: it captures much information on the community structure, and it can be used in an efficient hierarchical agglomerative algorithm that detects communities in a network. We designed such an algorithm which works in the worst case in time $\mathcal{O}(mn^2)$. In practice, real-world complex networks are sparse ($m = \mathcal{O}(n)$) and the height of the dendrogram is small ($H = \mathcal{O}(\log n)$); in this case the algorithm runs in $\mathcal{O}(n^2\log n)$. An implementation is provided at \cite{url_prog}.

Extensive experiments show that our method provides good results in various conditions (graph sizes, densities, and number of communities). We used such experiments to compare our algorithms to the main previously proposed ones. This direct comparision shows that our approach has a clear advantage in term of quality of the computed partition and presents the best tradeoff between quality and running time for large networks. It however has the limitation of needing quite a large amount of memory, which makes the Fast Modularity approach a relevant challenger of our method for very large graphs (million vertices).

Our method could be integrated in a multi-scale visualization tool for large networks, and it may be relevant for the computation of \emph{overlapping} communities (which often occurs in real-world cases and on which very few has been done until now \cite{overlapping}). We consider these two points as promising directions for further work. Finally, we pointed out that the method is directly usable for {\em weighted} networks. For directed ones (like the important case of the web graph), on the contrary, the proofs we provided are not valid anymore, and random walks behave significantly differently. Therefore, we also consider the directed case as an interesting direction for further research.

\subsubsection*{Acknowledgments}

We first want to thank our colleagues who provided us an implementation of their algorithm. We also thank Annick Lesne and L.S. Shulman for useful conversation and Aaron Clauset and Cl\'emence Magnien for helpful comments on preliminary versions. This work has been supported in part by the PERSI ({\em Programme d'\'Etude des R\'eseaux Sociaux de l'Internet}) project and by the GAP ({\em Graphs, Algorithms and Probabilities}) project.

\bibliography{biblio}

\begin{thebibliography}{10}

\bibitem{albert:2002}
R.~Albert and A.-L. Barab\'asi.
\newblock Statistical mechanics of complex networks.
\newblock {\em Reviews of Modern Physics}, 74(1):47, 2002.

\bibitem{Albert:1999}
R\'eka Albert, Jeong Hawoong, and Barab\'asi Albert-L\'aszl\'o.
\newblock Diameter of the world wide web.
\newblock {\em Nature}, 401:130, 1999.

\bibitem{Aldenderfer:1984}
M.~S. Aldenderfer and R.~K. Blashfield.
\newblock {\em Cluster Analysis}.
\newblock Number 07-044 in Sage University Paper Series on Quantitative
  Applications in the Social Sciences. Sage, Beverly Hills, 1984.

\bibitem{book_Aldous}
D.~Aldous and J.~A. Fill.
\newblock {\em Reversible Markov Chains and Random Walks on Graphs}, chapter~2.
\newblock Forthcoming book,
  http://www.stat.berkeley.edu/users/aldous/RWG/book.html.

\bibitem{Bagrow_Bollt:2004}
Jim Bagrow and Erik Bollt.
\newblock A local method for detecting communities.
\newblock {\em Physical Review E}, 2005 (to appear).

\bibitem{cosmoweb}
T.~Bennouas, M.~Bouklit, and F.~de~Montgolfier.
\newblock Un mod\`ele gravitationnel du web.
\newblock In {\em 5\`eme Rencontres Francophones sur les aspects Algorithmiques
  des T\'el\'ecommunications (Algotel)}, Banyuls (France), 2003.

\bibitem{Clauset:2005}
Aaron Clauset.
\newblock Finding local community structure in networks.
\newblock {\em Physical Review E}, 72:026132, 2005.

\bibitem{Clauset_Newman:2004}
Aaron Clauset, M.~E.~J. Newman, and Cristopher Moore.
\newblock Finding community structure in very large networks.
\newblock {\em Physical Review E}, 70(6):066111, 2004.

\bibitem{LdaFCosta:2004}
Luciano da~Fontoura~Costa.
\newblock Hub-based community finding.
\newblock {\em arXiv:cond-mat/0405022}, 2004.

\bibitem{Donetti:2004}
L.~Donetti and M.~A. Mu\~{n}oz.
\newblock Detecting network communities: a new systematic and efficient
  algorithm.
\newblock {\em Journal of Statistical Mechanics}, 2004(10):10012, 2004.

\bibitem{Donetti:2005}
L.~Donetti and M.~A. Mu\~{n}oz.
\newblock Improved spectral algorithm for the detection of network communities.
\newblock {\em arXiv:physics/0504059}, 2005.

\bibitem{Dorogovtsev:2003}
S.N. Dorogovtsev and J.F.F. Mendes.
\newblock {\em Evolution of Networks: From Biological Nets to the Internet and
  WWW}.
\newblock Oxford University Press, Oxford, 2003.

\bibitem{Drineas:2004}
P.~Drineas, A.~Frieze, R.~Kannan, S.~Vempala, and V.~Vinay.
\newblock Clustering large graphs via the singular value decomposition.
\newblock {\em Machine Learning}, 56(1-3):9--33, 2004.

\bibitem{Duch_Arenas:2005}
Jordi Duch and Alex Arenas.
\newblock Community detection in complex networks using extremal optimization.
\newblock {\em arXiv:cond-mat/0501368}, 2005.

\bibitem{Everitt:2001}
B.~S. Everitt, S.~Landau, and M.~Leese.
\newblock {\em Cluster Analysis}.
\newblock Hodder Arnold, London, $4^{th}$ edition, 2001.

\bibitem{Fernandez_de_la_Vega:2003}
W.~Fernandez de~la Vega, Marek Karpinski, Claire Kenyon, and Yuval Rabani.
\newblock Approximation schemes for clustering problems.
\newblock In {\em Proceedings of the thirty-fifth annual ACM Symposium on
  Theory of computing, STOC}, pages 50--58. ACM Press, 2003.

\bibitem{Fiedler:1973}
M.~Fiedler.
\newblock Algebraic connectivity of graphs.
\newblock {\em Czechoslovak Math. J.}, 23:298--305, 1973.

\bibitem{Flake:2002}
G.~W. Flake, S.~Lawrence, C.~L. Giles, and F.~M. Coetzee.
\newblock Self-organization and identification of web communities.
\newblock {\em Computer}, 35(3):66--71, 2002.

\bibitem{Fortunato:2004}
Santo Fortunato, Vito Latora, and Massimo Marchiori.
\newblock Method to find community structures based on information centrality.
\newblock {\em Physical Review E}, 70(5):056104, 2004.

\bibitem{Fouss}
F.~Fouss, A.~Pirotte, and M.~Saerens.
\newblock A novel way of computing dissimilarities between nodes of a graph,
  with application to collaborative filtering.
\newblock In {\em Workshop on Statistical Approaches for Web Mining (SAWM)},
  pages 26--37, Pisa, 2004.

\bibitem{Gaume:2004}
B.~Gaume.
\newblock Balades al\'eatoires dans les petits mondes lexicaux.
\newblock {\em I3 Information Interaction Intelligence}, 4(2), 2004.

\bibitem{Gaveau:1999}
B.~Gaveau, A.~Lesne, and L.~S. Schulman.
\newblock Spectral signatures of hierarchical relaxation.
\newblock {\em Physics Letters A}, 258(4-6):222--228, July 1999.

\bibitem{Girvan_Newman:2002}
M.~Girvan and M.~E.~J. Newman.
\newblock Community structure in social and biological networks.
\newblock {\em PNAS}, 99(12):7821--7826, 2002.

\bibitem{Magoni}
Micka\"el Hoerdt and Damien Magoni.
\newblock Completeness of the internet core topology collected by a fast
  mapping software.
\newblock In {\em Proceedings of the 11th International Conference on Software,
  Telecommunications and Computer Networks}, pages 257--261, Split, Croatia,
  October 2003.

\bibitem{Hubert_Arabie}
L.~Hubert and P.~Arabie.
\newblock Comparing partitions.
\newblock {\em Journal of Classification}, 2:193--218, 1985.

\bibitem{Jambu}
M.~Jambu and Lebeaux M.-O.
\newblock {\em Cluster analysis and data analysis}.
\newblock North Holland Publishing, 1983.

\bibitem{protein}
Hawoong Jeong, Sean Mason, Albert-L\'aszl\'o Barab\'asi, and Zolt\'an~N.
  Oltvai.
\newblock Centrality and lethality of protein networks.
\newblock {\em Nature}, 411:41--42, 2001.

\bibitem{Kernighan:1970}
B.~W. Kernighan and S.~Lin.
\newblock An efficient heuristic procedure for partitioning graphs.
\newblock {\em Bell System Technical Journal}, 49(2):291--308, 1970.

\bibitem{kleinberg:2001}
Jon Kleinberg and Steve Lawrence.
\newblock The structure of the web.
\newblock {\em Science}, 294(5548):1849--1850, 2001.

\bibitem{Lovasz_random_walks}
L.~Lov{\'a}sz.
\newblock Random walks on graphs: a survey.
\newblock In {\em Combinatorics, Paul Erd\H os is eighty, Vol.\ 2 (Keszthely,
  1993)}, volume~2 of {\em Bolyai Soc. Math. Stud.}, pages 353--397. J\'anos
  Bolyai Math. Soc., Budapest, 1996.

\bibitem{Newman:2003}
M.~E.~J. Newman.
\newblock The structure and function of complex networks.
\newblock {\em SIAM REVIEW}, 45:167, 2003.

\bibitem{Newman:2004}
M.~E.~J. Newman.
\newblock Fast algorithm for detecting community structure in networks.
\newblock {\em Physical Review E}, 69(6):066133, 2004.

\bibitem{Newman_Girvan:2004}
M.~E.~J. Newman and M.~Girvan.
\newblock Finding and evaluating community structure in networks.
\newblock {\em Physical Review E}, 69(2):026113, 2004.

\bibitem{overlapping}
Gergely Palla, Imre Derenyi, Illes Farkas, and Tamas Vicsek.
\newblock Uncovering the overlapping community structure of complex networks in
  nature and society.
\newblock {\em Nature}, 435:814--818, 2005.

\bibitem{Pothen:1990}
A.~Pothen, H.~D. Simon, and K.-P. Liou.
\newblock Partitioning sparse matrices with eigenvectors of graphs.
\newblock {\em SIAM J. Matrix Anal. Appl.}, 11(3):430--452, 1990.

\bibitem{Radicchi_Filippo:2004}
F.~Radicchi, C.~Castellano, F.~Cecconi, V.~Loreto, and D.~Parisi.
\newblock {Defining and identifying communities in networks}.
\newblock {\em PNAS}, 101(9):2658--2663, 2004.

\bibitem{Rand}
W.M. Rand.
\newblock Objective criteria for the evaluation of clustering methods.
\newblock {\em Journal of the American Statistical association}, 66:846--850,
  1971.

\bibitem{Ravasz:2002}
E.~Ravasz, A.~L. Somera, D.~A. Mongru, Z.~N. Oltvai, and A.-L. Barab\'asi.
\newblock {Hierarchical Organization of Modularity in Metabolic Networks}.
\newblock {\em Science}, 297(5586):1551--1555, 2002.

\bibitem{Reichardt_Bornholdt:2004}
J\"org Reichardt and Stefan Bornholdt.
\newblock Detecting fuzzy community structures in complex networks with a potts
  model.
\newblock {\em Physical Review Letters}, 93:218701, 2004.

\bibitem{Schulman:2001}
L.~S. Schulman and B.~Gaveau.
\newblock Coarse grains: The emergence of space and order.
\newblock {\em Foundations of Physics}, 31(4):713--731, April 2001.

\bibitem{Simonsen:2004}
I.~Simonsen, K.~Astrup Eriksen, S.~Maslov, and K.~Sneppen.
\newblock Diffusion on complex networks: a way to probe their large-scale
  topological structures.
\newblock {\em Physica A: Statistical Mechanics and its Applications},
  336(1-2):163--173, May 2004.

\bibitem{Strogatz:2001}
S.~H. Strogatz.
\newblock Exploring complex networks.
\newblock {\em Nature}, 410:268--276, March 2001.

\bibitem{mcl}
Stijn van Dongen.
\newblock {\em Graph Clustering by Flow Simulation}.
\newblock PhD thesis, University of Utrecht, May 2000.

\bibitem{Ward:1963}
J.~H. Ward.
\newblock Hierarchical grouping to optimize an objective function.
\newblock {\em Journal of the American Statistical Association},
  58(301):236--244, 1963.

\bibitem{wasserman94socialnetwork}
S.~Wasserman and K.~Faust.
\newblock {\em Social network analysis}.
\newblock Cambridge University Press, Cambridge, 1994.

\bibitem{Wu_Huberman:2003}
Fang Wu and Bernardo~A. Huberman.
\newblock Finding communities in linear time: A physics approach.
\newblock {\em The European Physical Journal B}, 38:331--338, 2004.

\bibitem{Zachary}
Wayne~W. Zachary.
\newblock An information flow model for conflict and fission in small groups.
\newblock {\em Journal of Anthropological Research}, 33:452--473, 1977.

\bibitem{ZhouL04}
Haijun Zhou and Reinhard Lipowsky.
\newblock Network brownian motion: A new method to measure vertex-vertex
  proximity and to identify communities and subcommunities.
\newblock In {\em International Conference on Computational Science}, pages
  1062--1069, 2004.

\bibitem{url_prog}
\verb#http://liafa.jussieu.fr/~pons/#.

\bibitem{url_data_arxiv}
data set obtained from
  \verb#http://www.cs.cornell.edu/projects/kddcup/datasets.html#.

\end{thebibliography}
\bibliographystyle{plain}

\newpage
\appendix 

\end{document}